\documentclass[usenatbib,letters]{mn2e}
\usepackage{amsmath}
\usepackage{amssymb}
\usepackage{graphicx}
\usepackage{aas_macros}

\author[B. Farris et al.]{Brian D.~Farris$^{1,2}$, Paul~Duffell$^{1}$, Andrew I.~MacFadyen$^{1}$, and Zolt\'an~Haiman$^{2}$ \\
    $^1$Center for Cosmology and Particle Physics, Physics Department, New York University, New York, NY 10003, USA\\
$^2$Department of Astronomy, Columbia University, 550 West 120th Street, New York, NY 10027, USA} 

\title{Characteristic Signatures in the Thermal Emission from Accreting Binary Black Holes}

\begin{document}

\maketitle
\begin{abstract}
    We present the results of a calculation of the thermal spectrum from a 2D, moving mesh, high-accuracy, viscous hydrodynamical simulation of an accreting supermassive black hole binary. We include viscous heating, shock heating, and radiative cooling, evolving for longer than a viscous time so that we reach a quasi-steady accretion state. In agreement with previous work, we find that gas is efficiently stripped from the inner edge of the circumbinary disk and enters the cavity along accretion streams, which feed persistent ``mini-disks" surrounding each black hole. We also find that emission from the shock-heated mini-disks and accretion streams prevents any deficit in high-energy emission that may be expected inside the circumbinary cavity, and instead leads to a characteristic brightening of the spectrum beginning in soft X-rays. 
\end{abstract}

\section{Introduction}
Supermassive black hole (SMBHs) binaries have gained significant attention as potential sources of both gravitational and electromagnetic radiation. It is currently believed that SMBHs with masses between $10^6$ and $10^9 M_{\sun}$ reside in nearly all nearby galaxy nuclei \citep{kormendy95,ferrarese05}. Mergers of such galaxies are expected to give rise to bound SMBH binary systems in the merged galaxy remnant, following a phase of dynamical friction in which the SMBHs become gravitationally bound (see the recent review by \citealt{mayer13}).

It is expected that there is an abundance of dense gas in the nuclei of merged galaxies \citep{barnes92,springel05} and that this gas can form a circumbinary accretion disk \citep{artymowicz96,armitage02,milos05}. Such binaries may provide a unique opportunity to observe
electromagnetic signatures
originating from the interaction of the binary with the surrounding
accretion disk. The gravitational radiation originating from the SMBH
binary inspiral should also be detectable by Pulsar Timing Arrays
(PTAs) \citep{hobbs10,kocsis11,tanaka12,lommen12,sesana12} or by a space interferometer such
as eLISA \citep{amaroseoane13}, provided the binary has maintained a circumbinary disk past the decoupling epoch \citep{barausse08,noble12, kocsis12b}

A number of ``dual" systems, in which two SMBHs occupy the same galaxy, but are too widely separated to be gravitationally bound, have been observed \citep{komossa03, comerford13,liu13,woo14}, as well as several candidate binary systems (see e.g. \citet{liu14} and references therein).
Proposed electromagnetic signatures of such binaries include spatially resolving two AGN-like point-sources, identifying double-peaked broad emission lines, spatial structures in radio jets, and characteristic time variability in quasar emission. 

In this paper we focus on characteristic changes to the thermal continuum emission brought about by the presence of a binary. Such signatures have been discussed previously \citep{liu03,milos05,hayasaki08,lodato09,shapiro10,liu10,tanaka12,gultekin12,kocsis12b,tanaka13,roedig14}, but we present the first calculation of a SMBH thermal spectrum directly from simulation results. We find that our spectra differ qualitatively from prior predictions due to the large amount of emission from hot gas residing in minidisks surrounding each BH, as well as in the accretion streams which penetrate the cavity.  This gas causes a surplus, rather than a deficit, in the high-energy regions of the composite spectrum.

\section{Methods}
Our initial disk configurations consist of the ``middle region" Shakura-Sunyaev solution for a steady-state, geometrically thin, optically thick accretion disk, assuming a gas pressure dominated fluid with electron scattering as the dominant opacity. We modify this solution by exponentially reducing the density and pressure by a factor of $\mbox{exp}\left[-(r/r_0)^{-10}\right]$ to create a hollow cavity within $r \lesssim r_0 \equiv 2.5 a$, where $a$ is the binary separation. In the early stage of our simulations, the inner cavity wall migrates inward due to viscosity, accretion streams form, and a quasi-steady state is reached. The fluid evolves according the the 2D viscous Navier Stokes equations, assuming an $\alpha$-law viscosity prescription and accounting for the binary potential in the expression for the disk scale height $h$,
\begin{equation}
\nu = \alpha c_s h = \alpha \frac{P}{\Sigma}\left(\frac{G m_1}{r_1^3}+\frac{G m_2}{r_2^3}\right)^{-1/2} \ ,
\end{equation}
where $\Sigma$ is the surface density, $P$ is the vertically integrated pressure, $m_i$ is the mass of each BH, $r_i$ is the distance to each BH, and $\alpha$ is the viscosity parameter, which we choose to be $\alpha=0.1$.

The {\it DISCO} code \citep{duffell14} allows one the freedom to specify the motion of the computational cells. For this problem we choose a rotation profile which matches the nearly Keplerian fluid motion outside the cavity, while transitioning to uniform rotation at the binary orbital frequency inside the cavity. 

\begin{figure}
\centering  
\includegraphics[width=0.99\columnwidth]{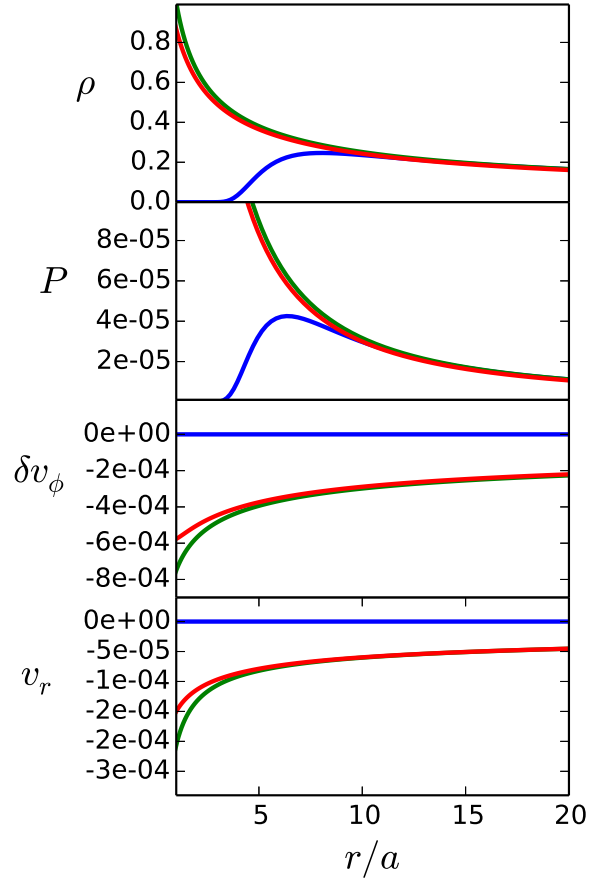}
\caption{Results of Shakura-Sunyaev solution. The initial condition, shown in blue, has an inner cavity, but  outside $r\gtrsim5a$, it matches the analytic solution for a gas-pressure dominated Shakura-Sunyaev disk with scale height $h/r = 0.03$ around a single BH with electron scattering as the dominant opacity. The approximate (to first order in $1/\mathcal{M}_a$)
 analytic solution for density $\rho$, pressure $P$, deviation from Keplerian azimuthal velocity $\delta v^{\phi} \equiv v^{\phi} - v^{\phi}_{kep}$ and radial velocity $v_r$ is shown in green, and the relaxed simulation results are shown in red. }
\label{fig:SStest}
\end{figure}

Our numerical methods are similar to those described in \citet{farris14}, with several important differences. We have replaced the isothermal prescription with a full evolution of the energy equation, assuming a $\Gamma$-law equation of state of the form $P=(\Gamma - 1)\epsilon$, where $\epsilon$ is the internal energy density, and the adiabatic index is set to $\Gamma=5/3$. We have added the appropriate radiative cooling and viscous heating terms to the energy equation. The cooling rate for an optically thick, geometrically thin disk is $q_{cool} = 4/3\tau^{-1} T^4$, where $T$ is the mid-plane temperature,
and $\tau$ is the optical depth for electron scattering ($\tau=\Sigma\sigma_T/m_p$). For a Shakura-Sunyaev disk around a single BH, this is in balance with the local viscous heating rate, $q_{vis} = 9/8 \alpha P \Omega$ \citep{frank02}. We therefore parametrize our simulations by choosing the Mach number at $r=a$, $\mathcal{M}_a \equiv (GM/a)^{1/2}/\sqrt{P(a)/\Sigma(a)}$, and scaling the cooling rate accordingly,
\begin{equation}
    q_{cool} = \frac{9}{8} \alpha \Omega(a) P(a) \mathcal{M}_a^4\left(\frac{P}{\Sigma v_a^2}\right)^4  \left(\frac{\Sigma}{\Sigma(a)}\right)^{-1} \ .
\end{equation}
For the simulations in this paper, we set $\mathcal{M}_a = 32$.
Details of our implementation of viscous heating in the energy equation are straightforward, and will be described in detail in an upcoming paper. 
With these improvements, we can self-consistently determine the temperature and emission everywhere in the disk due to both viscous heating and shock heating. We have tested that the code can reproduce the analytic steady-state Shakura-Sunyaev solution around a single BH (see Fig.~\ref{fig:SStest}). This demonstrates our ability to accurately balance viscous heating and radiative cooling. 

\begin{figure}
\centering  
\includegraphics[width=0.99\columnwidth]{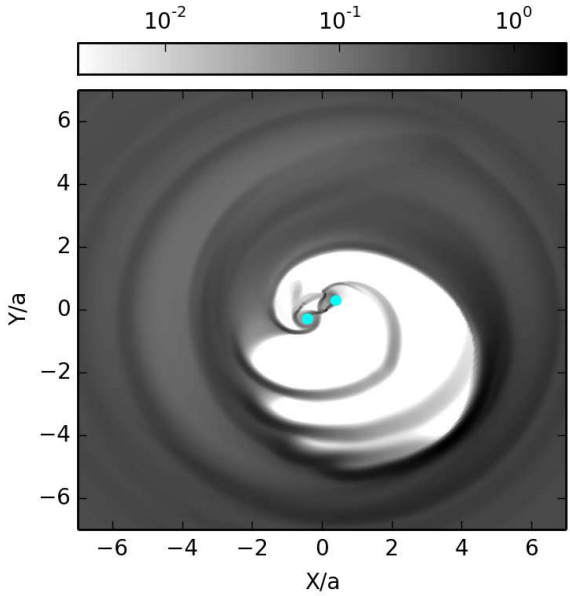}
\caption{Snapshot of surface density $\Sigma$ during quasi-steady state after $t \approx 2700 t_{bin} \approx t_{vis}$. Surface density is normalized by the maximum value at $t=0$ and plotted on a logarithmic scale in the inner $\pm 6a$. Orbital motion is in the counter-clockwise direction.}
\label{fig:density}
\end{figure}

\begin{figure}
\centering  
\includegraphics[width=0.99\columnwidth]{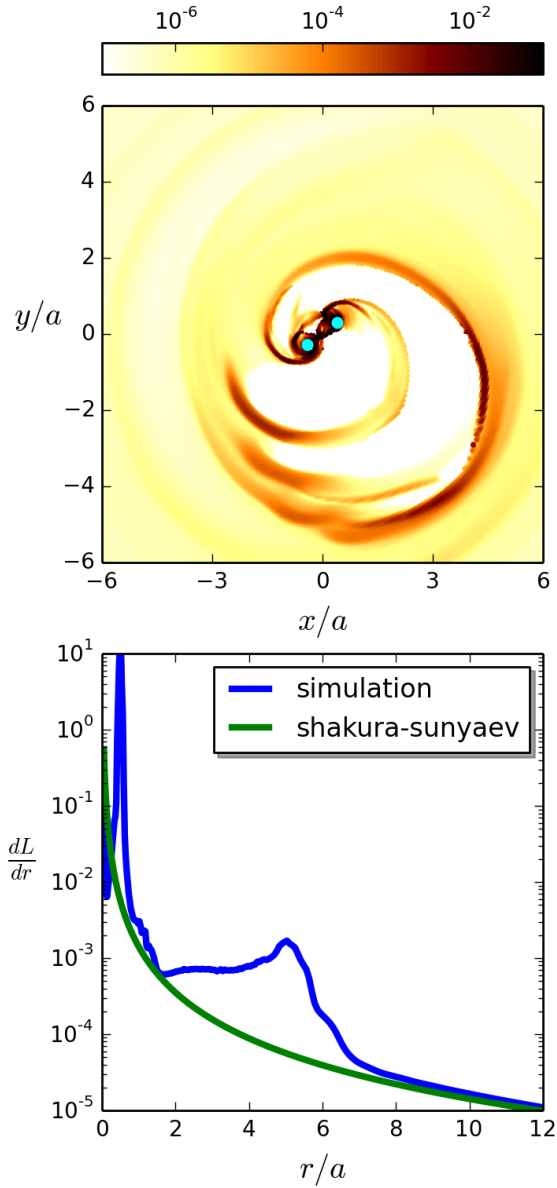}
\caption{{\em Top Panel}: Snapshot of surface brightness  $d L / dA$ during quasi-steady state after $t \approx 2700 t_{bin} \approx t_{vis}$, normalized by the value at $r=a$ for a steady-state disk around a single BH. {\em Bottom Panel}: Radial profile of azimuthally integrated surface brightness, $dL/dr$ time averaged over $\approx 25$ orbits is plotted in blue, reference profile for a steady-state disk around a single BH plotted in green for comparison.}
\label{fig:emission}
\end{figure}

In Figure~\ref{fig:density}, we plot a snapshot of the gas density near the binary at $t = 2700 t_{bin}$, after the disk has relaxed to a quasi-equilibrium state. We note that the cavity has grown lopsided, as observed in, e.g. \citet{macfadyen08,cuadra09,roedig12,shi12,noble12,farris14}. 
See \citet{farris14} and \citet{shi12} for a description of the mechanism driving the growth of this lopsidedness. 
The results are qualitatively similar to those of \citet{farris14}, with prominent streams entering the cavity and feeding persistent minidisks around each BH. 
By treating the radiative cooling of the disk self-consistently, we are able to calculate profiles of the electromagnetic emission from these disks. In Fig~\ref{fig:emission} (top panel) we plot a 2D profile of the bolometric luminosity per unit surface area. We note that the brightest component of the accretion flow is the hot minidisks, with significant emission coming from the shock-heated fluid in the accretion streams. We also plot the azimuthally integrated luminosity per annulus, $dL/dr$ and compare with the expected Shakura-Sunyaev profile for a steady-state disk around a single BH. While the enhancement near $r/a=0.5$ is expected due to the hot minidisks surrounding each BH, it is surprising that there is no dip in $dL/dr$ associated with the low density cavity, as has been widely assumed in the literature (see e.g. \citealt{milos05,tanakamenou10}). 
On the contrary, we find that the emission associated with the shock heating in the accretion streams is sufficient to bring the emission from within the cavity above that of a disk around a single BH.

\begin{figure}
\centering  
\includegraphics[width=0.99\columnwidth]{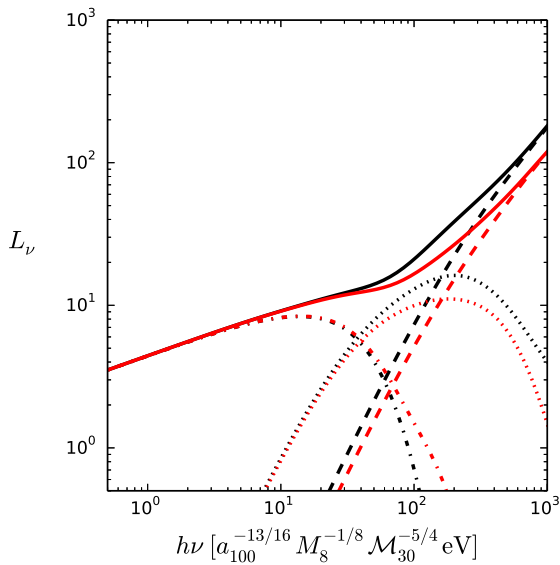}
\caption{Thermal spectra computed from simulation snapshot at $t \approx 2700 t_{bin} \approx t_{vis}$. Red curves are calculated from disk data $\approx 3 t_{bin}$ after that of black curves. The full spectra is represented by solid lines, the component arising only from minidisk regions within distance $d<0.5a$ of either BH is represented by dashed lines, the ``cavity" emission is represented by dotted lines, and the emission from the ``outer region" is represented by dashed-dotted lines.}
\label{fig:spectrum}
\end{figure}

\begin{figure}
\centering  
\includegraphics[width=0.99\columnwidth]{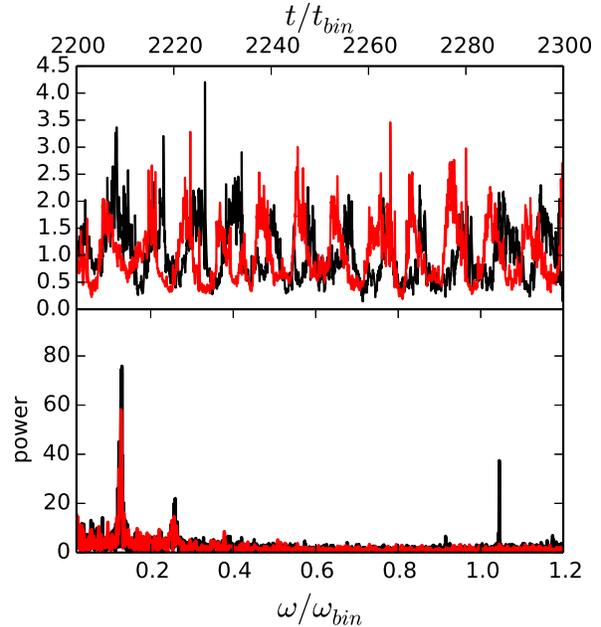}
\caption{The time variable accretion rate (black) onto the binary, and total luminosity (red), are plotted in the top panel over a window of 100 orbits, and the corresponding Lomb-Scargle periodograms, computed over $\sim 1000$ orbits, are plotted in the bottom panel.}
\label{fig:periodogram}
\end{figure}

For realistic circumbinary disks at cosmological distances, features inside the cavity are unlikely to be directly resolved observationally. However, features in the circumbinary disk profile can leave an imprint in the observed AGN spectrum. In an ordinary AGN disk, the gas temperature depends strongly on radius, and thus thermal emission from an annulus at $r$ contributes to the spectrum most strongly at $h \nu \approx k T(r)$. Consequently, it has been predicted that the missing emission from the circumbinary cavity can lead to a ``notch" in the thermal spectrum \citep{roedig14}. We test this prediction in our simulations by calculating the spectrum assuming thermal blackbody emission at each point in the disk,

\begin{equation}
    L_{\nu} = \int \frac{2 h \nu^3}{c^2\mbox{exp}\left(\frac{h \nu}{kT_{\rm{eff}}(r,\phi)}\right)-1}dA
\end{equation}
where the effective temperature $T_{\rm{eff}}(r,\phi) \equiv (q_{cool}(r,\phi) / \sigma)^{1/4}$, and $q_{cool}$ is the radiative cooling rate.
We find no noticeable ``notch" in the spectrum, as the low-frequency tail of the emission from the hot minidisks and the hot streams effectively washes out any defecits arising from the missing gas in the cavity.
A spectrum computed from snapshots separated by $\delta t \approx 3t_{bin}$ 
during the relaxed, quasi-steady state at $t \approx 2700 t_{bin}$ is shown in Figure~\ref{fig:spectrum}. We find a strong enhancement in emission at high frequencies due to the shock heated minidisks. Scaled to a $10^8 M_{\odot}$ binary  with a separation near decoupling at $a/M = 100$, we find that the enhancement is significant in soft and hard X-rays. This X-ray enhancement has been predicted by \citet{roedig14}, who estimated the characteristic frequency of minidisk ``hot spot" emission by estimating the amount of energy released due to shock heating when accretion streams impact the minidisks. While our enhancements occur at somewhat lower energy compared to their estimate of $\sim 100 \mbox{keV}$, this is likely due to our simplifying assumption of thermal blackbody emission throughout our domain, whereas they consider emission due to inverse Compton scattering. We also expect that spectral enhancements are likely to extend to lower frequencies when one considers emission from the hot streams within the cavity. We decompose the spectrum into a component emitted from within the minidisk region inside a distance $a/2$ of each black hole, a ``cavity" component emitted from outside the minidisks, but within a distance of $6a$ from the origin, and an ``outer" region emitted from outside $6a$. We plot spectra at two different times, separated by $\sim 3 t_{bin}$ in order to demonstrate the time variability of the spectrum. While the emission from the ``outer" region is steady in time, the emission from the cavity and the minidisks shifts to higher frequencies as a high density ``lump" falls on the binary, coinciding with a spike in $\dot{M}$. Such lumps have been noted in a number of previous simulations \citep{macfadyen08,roedig11,noble12,shi12,farris14}.
 In Figure~\ref{fig:periodogram}, we plot the accretion rate $\dot{M}$ and total luminosity $L$ as a function of time, as well as the respective Lomb-Scargle periodograms. We see that the accretion rate remains roughly proportional to the luminosity, as both vary by a factor of $\sim 4$. In both cases the dominant mode in the periodogram is at $\omega \approx 1/7 \omega_{bin}$, corresponding to the orbital frequency of the high density lump in the disk just outside the cavity.

%

\section{Discussion}
Accreting SMBH binaries may constitute rare sources with observable electromagnetic as well as gravitational radiation. In order to distinguish electromagnetic emission from binaries from that of ordinary AGN, it is important to identify observable signatures that characterize binary accretion. The Shakura-Sunyaev model has been successful in explaining the continuum spectra of ordinary AGN. Here we present a generalization of this model by computing the blackbody spectrum from the results of 2D simulations of binary accretion.

Prior work has focused on deficits in the spectrum which correspond to missing emission from the circumbinary cavity. We find that emission from the hot streams which penetrate the cavity, along with the low-frequency tail of the emission from the shock-heated minidisks is sufficient to mask such ``notches". Other proposed signatures which depend on a high-frequency cutoff associated with an empty cavity such as the ``Lyman edges" predicted by \citet{generozov14} are disfavored by our results as well (although this signature could become visible if the minidisks were radiatively inefficient).
We find that enhanced X-ray emission from the minidisks is a more prominent and robust signature. Such ``hot spots" have been previously predicted at somewhat higher frequencies using analytic arguments \citep{roedig14}, and were also predicted to arise from the tidally heated rim of the cavity \citep{lodato09,kocsis12b},
but we have provided the first self-consistent numerical simulations which demonstrate this feature. Instruments sensitive to X-ray emission from AGN such as {\it XMM-Newton}\footnote{http://sci.esa.int/xmm-newton/}, {\it NuSTAR}\footnote{http://www.nustar.caltech.edu/}, the upcoming {\it
eROSITA}\footnote{http://www.mpe.mpg.de/erosita/} all-sky survey, and the proposed {\it ATHENA}\footnote{http://www.the-athena-x-ray-observatory.eu/} X-ray observatory may be sensitive to these signatures. We also note that broad lines, which are powered by the hard radiation from the inner disk, may be enhanced by this emission, contrary to the unusually weak broad lines expected in the case of a dim cavity \citep{tanaka12}.

In future work, we intend to relax our black-body assumption and take into account radiative processes such as inverse Compton scattering that may be relevant. We also intend to include radiation pressure in our calculations, which may become particularly relevant in the shock-heated minidisks. In this paper, we have focused our attention on equal-mass binaries, but we intend to study the dependence of spectral features on binary mass-ratio. We also intend to determine the dependence of these features on disk thicknesses. For computational efficiency, we have performed our simulations in 2D, assuming an $\alpha$-law viscosity as a proxy for the viscosity arising from magneto-hydrodynamic (MHD) turbulence. In future work, we intend to perform full 3D MHD simulations using the {\it DISCO} code \citep{duffell14} in order to verify the validity of these approximations.

Resources supporting this work were provided by the NASA High-End Computing (HEC) Program through the NASA Advanced Supercomputing (NAS) Division at Ames Research Center and by the High Performance Computing resources at New York University Abu Dhabi. We acknowledge support from NASA grant NNX11AE05G (to ZH and AM). We are grateful to Daniel D'Orazio, Andrei Gruzinov and Aleksey Generozov for helpful comments and discussions. 

\bibliographystyle{mn2e}
\bibliography{refs}

\begin{thebibliography}{40}
\expandafter\ifx\csname natexlab\endcsname\relax\def\natexlab#1{#1}\fi

\bibitem[{{Amaro-Seoane} {et~al}\mbox{.}(2013){Amaro-Seoane}, {Aoudia},
  {Babak}, {Bin{\'e}truy}, {Berti}, {Boh{\'e}}, {Caprini}, {Colpi}, {Cornish},
  {Danzmann}, {Dufaux}, {Gair}, {Jennrich}, {Jetzer}, {Klein}, {Lang}, {Lobo},
  {Littenberg}, {McWilliams}, {Nelemans}, {Petiteau}, {Porter}, {Schutz},
  {Sesana}, {Stebbins}, {Sumner}, {Vallisneri}, {Vitale}, {Volonteri}, {Ward},
  \& {Wardell}}]{amaroseoane13}
{Amaro-Seoane} P. {et~al.}, 2013, GW Notes, Vol.~6, p.~4-110, 6, 4

\bibitem[{{Armitage} \& {Natarajan}(2002)}]{armitage02}
{Armitage} P.~J., {Natarajan} P., 2002, \apjl, 567, L9

\bibitem[{{Artymowicz} \& {Lubow}(1996)}]{artymowicz96}
{Artymowicz} P., {Lubow} S.~H., 1996, \apjl, 467, L77

\bibitem[{{Barausse} \& {Rezzolla}(2008)}]{barausse08}
{Barausse} E., {Rezzolla} L., 2008, \prd, 77, 104027

\bibitem[{{Barnes} \& {Hernquist}(1992)}]{barnes92}
{Barnes} J.~E., {Hernquist} L., 1992, \araa, 30, 705

\bibitem[{{Comerford} {et~al}\mbox{.}(2013){Comerford}, {Schluns}, {Greene}, \&
  {Cool}}]{comerford13}
{Comerford} J.~M., {Schluns} K., {Greene} J.~E., {Cool} R.~J., 2013, \apj, 777,
  64

\bibitem[{{Cuadra} {et~al}\mbox{.}(2009){Cuadra}, {Armitage}, {Alexander}, \&
  {Begelman}}]{cuadra09}
{Cuadra} J., {Armitage} P.~J., {Alexander} R.~D., {Begelman} M.~C., 2009,
  \mnras, 393, 1423

\bibitem[{{Duffell} \& {MacFadyen}(2014)}]{duffell14}
{Duffell} P., {MacFadyen} A.~I., 2014, in prep.

\bibitem[{{Farris} {et~al}\mbox{.}(2014){Farris}, {Duffell}, {MacFadyen}, \&
  {Haiman}}]{farris14}
{Farris} B.~D., {Duffell} P., {MacFadyen} A.~I., {Haiman} Z., 2014, \apj, 783,
  134

\bibitem[{{Ferrarese} \& {Ford}(2005)}]{ferrarese05}
{Ferrarese} L., {Ford} H., 2005, \ssr, 116, 523

\bibitem[{{Frank}, {King} \& {Raine}(2002){Frank}, {King}, \&
  {Raine}}]{frank02}
{Frank} J., {King} A., {Raine} D.~J., 2002, {Accretion Power in Astrophysics:
  Third Edition}

\bibitem[{{Generozov} \& {Haiman}(2014)}]{generozov14}
{Generozov} A., {Haiman} Z., 2014, ArXiv e-prints

\bibitem[{{G{\"u}ltekin} \& {Miller}(2012)}]{gultekin12}
{G{\"u}ltekin} K., {Miller} J.~M., 2012, \apj, 761, 90

\bibitem[{{Hayasaki}, {Mineshige} \& {Ho}(2008){Hayasaki}, {Mineshige}, \&
  {Ho}}]{hayasaki08}
{Hayasaki} K., {Mineshige} S., {Ho} L.~C., 2008, \apj, 682, 1134

\bibitem[{{Hobbs} {et~al}\mbox{.}(2010){Hobbs}, {Archibald}, {Arzoumanian},
  {Backer}, {Bailes}, {Bhat}, {Burgay}, {Burke-Spolaor}, {Champion}, {Cognard},
  {Coles}, {Cordes}, {Demorest}, {Desvignes}, {Ferdman}, {Finn}, {Freire},
  {Gonzalez}, {Hessels}, {Hotan}, {Janssen}, {Jenet}, {Jessner}, {Jordan},
  {Kaspi}, {Kramer}, {Kondratiev}, {Lazio}, {Lazaridis}, {Lee}, {Levin},
  {Lommen}, {Lorimer}, {Lynch}, {Lyne}, {Manchester}, {McLaughlin}, {Nice},
  {Oslowski}, {Pilia}, {Possenti}, {Purver}, {Ransom}, {Reynolds}, {Sanidas},
  {Sarkissian}, {Sesana}, {Shannon}, {Siemens}, {Stairs}, {Stappers},
  {Stinebring}, {Theureau}, {van Haasteren}, {van Straten}, {Verbiest},
  {Yardley}, \& {You}}]{hobbs10}
{Hobbs} G. {et~al.}, 2010, Classical and Quantum Gravity, 27, 084013

\bibitem[{{Kocsis}, {Haiman} \& {Loeb}(2012){Kocsis}, {Haiman}, \&
  {Loeb}}]{kocsis12b}
{Kocsis} B., {Haiman} Z., {Loeb} A., 2012, \mnras, 427, 2680

\bibitem[{{Kocsis} \& {Sesana}(2011)}]{kocsis11}
{Kocsis} B., {Sesana} A., 2011, \mnras, 411, 1467

\bibitem[{{Komossa} {et~al}\mbox{.}(2003){Komossa}, {Burwitz}, {Hasinger},
  {Predehl}, {Kaastra}, \& {Ikebe}}]{komossa03}
{Komossa} S., {Burwitz} V., {Hasinger} G., {Predehl} P., {Kaastra} J.~S.,
  {Ikebe} Y., 2003, \apjl, 582, L15

\bibitem[{{Kormendy} \& {Richstone}(1995)}]{kormendy95}
{Kormendy} J., {Richstone} D., 1995, \araa, 33, 581

\bibitem[{{Liu}, {Li} \& {Komossa}(2014){Liu}, {Li}, \& {Komossa}}]{liu14}
{Liu} F.~K., {Li} S., {Komossa} S., 2014, \apj, 786, 103

\bibitem[{{Liu}, {Wu} \& {Cao}(2003){Liu}, {Wu}, \& {Cao}}]{liu03}
{Liu} F.~K., {Wu} X.-B., {Cao} S.~L., 2003, \mnras, 340, 411

\bibitem[{{Liu} {et~al}\mbox{.}(2013){Liu}, {Civano}, {Shen}, {Green},
  {Greene}, \& {Strauss}}]{liu13}
{Liu} X., {Civano} F., {Shen} Y., {Green} P., {Greene} J.~E., {Strauss} M.~A.,
  2013, \apj, 762, 110

\bibitem[{{Liu} \& {Shapiro}(2010)}]{liu10}
{Liu} Y.~T., {Shapiro} S.~L., 2010, \prd, 82, 123011

\bibitem[{{Lodato} {et~al}\mbox{.}(2009){Lodato}, {Nayakshin}, {King}, \&
  {Pringle}}]{lodato09}
{Lodato} G., {Nayakshin} S., {King} A.~R., {Pringle} J.~E., 2009, \mnras, 398,
  1392

\bibitem[{{Lommen}(2012)}]{lommen12}
{Lommen} A.~N., 2012, Journal of Physics Conference Series, 363, 012029

\bibitem[{{MacFadyen} \& {Milosavljevi{\'c}}(2008)}]{macfadyen08}
{MacFadyen} A.~I., {Milosavljevi{\'c}} M., 2008, \apj, 672, 83

\bibitem[{{Mayer}(2013)}]{mayer13}
{Mayer} L., 2013, Classical and Quantum Gravity, 30, 244008

\bibitem[{{Milosavljevi{\'c}} \& {Phinney}(2005)}]{milos05}
{Milosavljevi{\'c}} M., {Phinney} E.~S., 2005, \apjl, 622, L93

\bibitem[{{Noble} {et~al}\mbox{.}(2012){Noble}, {Mundim}, {Nakano}, {Krolik},
  {Campanelli}, {Zlochower}, \& {Yunes}}]{noble12}
{Noble} S.~C., {Mundim} B.~C., {Nakano} H., {Krolik} J.~H., {Campanelli} M.,
  {Zlochower} Y., {Yunes} N., 2012, \apj, 755, 51

\bibitem[{{Roedig} {et~al}\mbox{.}(2011){Roedig}, {Dotti}, {Sesana}, {Cuadra},
  \& {Colpi}}]{roedig11}
{Roedig} C., {Dotti} M., {Sesana} A., {Cuadra} J., {Colpi} M., 2011, \mnras,
  415, 3033

\bibitem[{{Roedig}, {Krolik} \& {Miller}(2014){Roedig}, {Krolik}, \&
  {Miller}}]{roedig14}
{Roedig} C., {Krolik} J.~H., {Miller} M.~C., 2014, \apj, 785, 115

\bibitem[{{Roedig} {et~al}\mbox{.}(2012){Roedig}, {Sesana}, {Dotti}, {Cuadra},
  {Amaro-Seoane}, \& {Haardt}}]{roedig12}
{Roedig} C., {Sesana} A., {Dotti} M., {Cuadra} J., {Amaro-Seoane} P., {Haardt}
  F., 2012, \aap, 545, A127

\bibitem[{{Sesana} {et~al}\mbox{.}(2012){Sesana}, {Roedig}, {Reynolds}, \&
  {Dotti}}]{sesana12}
{Sesana} A., {Roedig} C., {Reynolds} M.~T., {Dotti} M., 2012, \mnras, 420, 860

\bibitem[{{Shapiro}(2010)}]{shapiro10}
{Shapiro} S.~L., 2010, \prd, 81, 024019

\bibitem[{{Shi} {et~al}\mbox{.}(2012){Shi}, {Krolik}, {Lubow}, \&
  {Hawley}}]{shi12}
{Shi} J.-M., {Krolik} J.~H., {Lubow} S.~H., {Hawley} J.~F., 2012, \apj, 749,
  118

\bibitem[{{Springel}, {Di Matteo} \& {Hernquist}(2005){Springel}, {Di Matteo},
  \& {Hernquist}}]{springel05}
{Springel} V., {Di Matteo} T., {Hernquist} L., 2005, \apjl, 620, L79

\bibitem[{{Tanaka} \& {Menou}(2010)}]{tanakamenou10}
{Tanaka} T., {Menou} K., 2010, \apj, 714, 404

\bibitem[{{Tanaka}, {Menou} \& {Haiman}(2012){Tanaka}, {Menou}, \&
  {Haiman}}]{tanaka12}
{Tanaka} T., {Menou} K., {Haiman} Z., 2012, \mnras, 420, 705

\bibitem[{{Tanaka}(2013)}]{tanaka13}
{Tanaka} T.~L., 2013, ArXiv e-prints

\bibitem[{{Woo} {et~al}\mbox{.}(2014){Woo}, {Cho}, {Husemann}, {Komossa},
  {Park}, \& {Bennert}}]{woo14}
{Woo} J.-H., {Cho} H., {Husemann} B., {Komossa} S., {Park} D., {Bennert} V.~N.,
  2014, \mnras, 437, 32

\end{thebibliography}

\end{document}